BIOLOGICAL SCIENCES: Biophysics and Computational Biology

# The impact of deleterious passenger mutations on cancer progression

Christopher D McFarland[1], Gregory V Kryukov[2, 3], Shamil Sunyaev[1, 2, 3], and Leonid A Mirny[1, 2, 4, 5,*]

[1] Harvard University, Graduate Program in Biophysics, Boston, MA 02115
[2] The Broad Institute of MIT and Harvard, Cambridge, MA 02139
[3] Division of Genetics, Brigham and Women's Hospital, Harvard Medical School, Boston, MA 02115
[4] Institute for Medical Engineering and Science, Massachusetts Institute of Technology
[5] Department of Physics, Massachusetts Institute of Technology, Cambridge, MA 02129

* Corresponding author:
Leonid A Mirny,
E25-526, MIT, 77 Mass ave, Cambridge, MA 02129,
Tel: 617-452-4862, E-mail: leonid@mit.edu

Cancer progression is driven by a small number of genetic alterations accumulating in a neoplasm. These few driver alterations reside in a cancer genome alongside tens of thousands of other mutations that are widely believed to have no role in cancer and termed passengers. Many passengers, however, fall within protein coding genes and other functional elements and can possibly have deleterious effects on cancer cells. Here we investigate a potential of mildly deleterious passengers to accumulate and alter the course of neoplastic progression. Our approach combines evolutionary simulations of cancer progression with the analysis of cancer sequencing data. In our simulations, individual cells stochastically divide, acquire advantageous driver and deleterious passenger mutations, or die. Surprisingly, despite selection against them, passengers accumulate and largely evade selection during progression. Although individually weak, the collective burden of passengers alters the course of progression leading to several phenomena observed in oncology that cannot be explained by a traditional driver-centric view. We tested predictions of the model using cancer genomic data. We find that many passenger mutations are likely to be damaging and that, in agreement with the model, they have largely evaded purifying selection. Finally, we used our model to explore cancer treatments that exploit the load of passengers by either 1) increasing the mutation rate; or 2) exacerbating their deleterious effects. While both approaches lead to cancer regression, the later leads to less frequent relapse. Our results suggest a new framework for understanding cancer progression as a balance of driver and passenger mutations.

# Introduction

Recent advances in sequencing and genotyping of cancer tissues at a genome level have found that individual cancers contain tens of thousands of somatic alterations (1-4). These alterations encompass many genetic alterations, such as single-nucleotide substitutions, insertions, deletions, rearrangements, loss of heterozygosity events (LOHs), copy number alterations, and whole chromosome duplications/deletions (1); epigenetic alterations (5); and other inheritable changes in cell state. It is generally believed that only a few (2-15) of these alterations cause the cancer phenotype, called *driver alterations* or simply *drivers*, while the overwhelming majority of events in cancer are believed to have non-significant phenotypes and are called *passenger alterations* or simply *passengers*. Drivers confer advantageous phenotypes to neoplastic cells (i.e. phenotypes that allow cells in the population to proliferate further), which is inferred by their alteration of cancer-related pathways; frequent occurrence at the same genes, loci, or pathways in different patients (3, 4, 6); and by the structure of cancer incidence rates (7). Because driver events are so critical to cancer progression, their discovery has been the primary goal of genome-wide cancer sequencing (8).

Conversely, little attention has been paid to passengers that are widely believed to have no impact on cancer progression (6, 9). These alterations are assumed to be phenotypically neutral in cancer cells because they are non-recurrent and are dispersed across a cancer genome (8, 10); however, their phenotype has never been systematically tested. Nevertheless, passengers constitute the vast majority of observed somatic alterations in cancer (**Table 2**) (4). If passengers arise as random alterations, then many can be deleterious to cells (11-13) via a variety of mechanisms, e.g. by disrupting housekeeping genes, inducing proteotoxic stress, or provoking an immune response. While highly deleterious passengers are weeded out by purifying selection, mildly deleterious passengers also experiencing purifying selection could nevertheless accumulate by mutation-selection balance, ratcheting and similar mechanisms studied in population genetics (13). Although individual passengers exert small effects, a large number of accumulated passengers can collectively be significant enough to alter the course of cancer progression.

Here we investigate the potential role of deleterious passenger alterations in cancer progression and examine their potential as a new therapeutic target. First, using a stochastic evolutionary model, where random passengers can arise alongside drivers in cancer cells, we find that mildly deleterious mutations can evade purifying selection and accumulate in cancer. We find that the accumulation of individually weak passengers nevertheless alters the dynamics of cancer progression and may explain several clinical phenomena, such as: slow progression, long periods of dormancy, the prevalence of small subclinical cancers, spontaneous regression, and genetic heterogeneity that cannot be easily explained without considering deleterious passengers. In contrast to the current paradigm of driver-centric cancer progression, our analyses demonstrate that progression depends on the balance of drivers and passengers. Second, we test a prediction of the model by analyzing somatic mutations sequenced in cancers. This analysis shows that, in agreement with the model, individual passengers are likely to be damaging to cells and that they have largely evaded purifying selection. Lastly, we use our model to explore two possible therapeutic approaches that target passengers and find that increasing either the mutation rate or the deleterious effect of passengers lead to rapid cancer meltdown. Since passengers may be deleterious via proteotoxicity (14, 15), loss of function (16), or by provoking an immune response (17), their effect can be aggravated by targeting pathways that buffer the effects of mutations, e.g. Unfolded Protein Response (UPR) pathways. We present and discuss clinical and biological evidence that supports an important role of passenger alterations in cancer.

## Results and Discussion

**A stochastic model of neoplastic progression incorporating passenger mutations.**

Mathematical modeling of cancer has a deep history (see (18) for review). However, existing evolutionary models have several limitations: many considered a population of a constant or externally controlled size (19, 20), which does not depend on the absolute fitness of the cells carrying specific mutations. Other models study exponentially growing cancer populations (7, 20, 21), while logistic-like behavior has been observed in cancer (22). Most importantly, the vast majority of cancer models (with the exception of (9), see below) neglected the effects of passenger alterations.

In our model individual cells divide or die stochastically, potentially acquiring driver and passenger alterations during division, and the population size can increase or shrink depending on the birth and death of individual cells (**Fig. 1A**). Most generally, the birth and death rates of cells depend on the effects of accumulated drivers and passengers, and the environment. Assuming that all drivers and passengers possess equal fitness advantage or disadvantage the birth and death rates $B(d,p,N)$ and $D(d,p,N)$ of each cell depend on the number of drivers $d$, the number of accumulated passengers $p$, and the total hyperplasia or population size $N$. Driver mutations increase the population size by either increasing the birth rate, e.g. an activating mutation in *KRAS*, or by decreasing the death rate, e.g. *p53* knockdown that diminishes contact inhibition and apoptosis (23). While specific drivers and passengers have differing effects on the birth and death rates, we find that aggregating the effects of mutations into the birth rate, and placing the effects of population size into the death rate does not alter population dynamics from models where mutational effects are split between the two (**Fig. S1**). Thus, we use

$$B(d,p) = \frac{(1+s_d)^d}{(1+s_p)^p}, \qquad D(N) = \frac{N}{K}$$

where $s_d$ is the fitness advantage (selection coefficient) of a driver, $s_p$ is the fitness disadvantage conferred by a passenger, and $K$ is the carrying capacity of the initial population, reflecting the effect of the tumor micro-environment. This form assumes multiplicative epistasis, but is equivalent to first order to other possible forms (e.g. $B(p,d) = (1+s_d)^d(1-s_p)^p$) (see supplement) such that it exhibits qualitatively similar behavior (**Fig. S1**). Our formalism extends previous models of neoplastic progression where only drivers affect growth rates (20, 24, 25). The death rate creates a logistic form of growth that reflects the effect of the tumor micro-environment and competition for recourses similar to previous neoplastic (24) and ecological (26) models. The linear dependence of $D(N)$ on $N$ is a coarse approximation of contact inhibition and the competition between cells for space and resources (e.g. due to a limited crypt size). In later simulations, we grew cancers to macroscopic size using $D(N) = log(1 + N/K)$. For small $N/K$ this reduces to the linear model above, but for large $N/K$ this recapitulates Gompertzian growth observed experimentally for large tumors (27). Tumors transition to Gompertzian growth perhaps because as the hyperplasia matures, it must overcome, via driver alterations, additional constraints on population size, such as: homeostatic pressure, hypoxia, competition for vascularization, immune and inflammatory responses, limited paracrine signaling, etc.

Cells acquire mutations with a constant rate $\mu$ and total driver/passenger mutation rates: $\mu T_{d/p}$, where $T_{d/p}$ are the mutation target sizes (in nucleotides) for drivers/passengers. While driver and passenger alterations can take many forms, we parameterized our model using single nucleotide substitution data, as these mutations have been more thoroughly quantified in cancer.

We model cancer progression as a stochastic system of various events with defined reaction rates (mutations, birth and death processes for each cell) and use a standard Gillespie algorithm (28). The system is fully defined by the following parameters: $s_p$, $s_d$, $\mu T_p$, $\mu T_d$, and $K$ that are estimated below.

Unlike previous models (19, 20), population size in our model varies stochastically with non-discrete generations. Like previous models (9), we consider 'death' to be any process that prevents a cell from replicating indefinitely, i.e. necrosis, apoptosis, senescence, or differentiation. Thus $N$ represents the population of cells that are capable of infinite division and of carrying the (epi)genetic information in cancer. Because we explored the initial population size across two orders of magnitude, our model should apply equally well to tumor sub-types dominated by only a small cohort of cancer stem cells, and sub-types where progenitor cells can spawn cancer (29). Our model lacks asymmetric cell divisions and ignores the spatial structure of cancer; both aspects presumably lead to an effectively smaller population size (30), which leads to a faster rate of passenger fixation (**Fig. S2**). Lastly, we included only passenger loci that, when mutated, have a mildly deleterious effect. Analytical treatment of our framework and more sophisticated models will be considered elsewhere.

**Estimating and exploring parameters of the model**

To account for the extensive heterogeneity of cancer between and within sub-types, and to account for our limited quantitative knowledge of this process, we varied the values of all parameters by 2-3 orders of magnitude. The ranges we explored centered about values derived from the literature (**Table 1**). We found that deleterious passenger mutations accumulate under this broad range of conditions (**Fig. S2**). The effect of each driver was assumed to be very significant ($s_d \approx 0.1$, i.e. individual drivers increase the growth rate by 10%) because previous studies found this rate to be congruent with the time to cancer onset (20). Simulations where this parameter was varied in the range of 0.001 to 1, or where drivers conferred a Gaussian or exponential distribution of fitness advantages did not differ qualitatively from our fixed-effect model (**Fig S1**). Recent estimates of the effects of near-neutral germ line mutations in humans (31) as well as randomly introduced mutations in yeast (15), suggest the disadvantage conferred by a passenger is very weak ($s_p \approx 0.001$, range $10^{-4} - 10^{-1}$).

The target size for driver mutations $T_d$, represents all relevant sites in all genes associated with cancer development (i.e. mutational hot spots in oncogenes and tumor suppressors). Our estimate, $T_d \approx 700$ positions per genome (70 genes (4) × approximately 10 activating mutations per gene; range 70-7,000) has been used previously in simulations (20), is projected from current sequencing data (4), and is approximately equal in number to the 571 loci with observed recurrent mutations in colon cancer (32). While the number of potentially inactivating mutations for any tumor suppressors is certainly larger than 10 per gene, and mutational effects should remain silent until a LOH event, for parsimony we assume that these nuances collectively result in an effective target size for a tumor suppressor that is still approximately 10. Because the number of observed mutations in major oncogenes is of the same order as the observed number of mutations in major tumor suppressors (32), we assume this approximation is reasonable.

Our estimate for the target size of potentially deleterious passengers represents ~5000 non-cancer related genes expressed in cancer, times ~1000 non-synonymous and non-neutral loci per gene (i.e $T_p \approx$ 5,000,000). This quantity is more conservative than previous estimates of 10,000,000 (33) and much greater than the target size for drivers. While we focus on non-synonymous passengers, many more sites in the genome are non-coding, which lead to $10^4 - 10^5$ mutations per cancer genome (2, 34). We assume these mutations have no effect on progression. The mutation rate ($\mu \approx 10^{-8}$ *nucleotide$^{-1}$ ·division$^{-}$*

[1], range $10^{-10} - 10^{-6}$) approximates cells that have acquired a mutator phenotype (35), while the initial carrying capacity ($K \approx 1,000$; range $100 - 10,000$) was estimated from the observed size of hyperplasias within a mouse colonic crypt 2 weeks following an initiating *APC* deletion (36).

**Mildly deleterious passengers fixate and influence the course of cancer progression**

**Figure 1B** presents typical population trajectories of a hyperplasia, beginning at the first activating mutation. First, it is evident that each trajectory consists of intervals of rapid growth and gradual decline. Acquisition of each new driver leads to a clonal expansion of the subpopulation carrying this driver, causing short periods of growth. Growth stops when the effect of the driver is balanced by the carrying capacity of the environment. While the population is waiting for the next driver to arise, passengers steadily accumulate leading to gradual decline of population size. These two processes lead to the sawtooth patterns of individual trajectories (**Fig 2A**).

Second, the trajectories exhibit two possible fates: rapid growth or gradual decline (regression), often after a period of dormancy (**Fig 1B**). These two alternative fates emerge due to the tug-of-war between the infrequent arrival of large-effect drivers, and frequent but mildly deleterious passengers. Simulations show that the fate of the population depends on its ability to reach a critical size: hyperplasias larger than the critical size are likely to progress, while smaller ones are likely to regress (**Fig S3**). This is because a larger population has a greater chance of acquiring the next driver in *any* of its cells—leading to more frequent clonal expansions (SI; analytical treatment). Dormancy and spontaneous regression observed in our model have also been observed experimentally (37) and do not occur in simulations lacking deleterious passengers (**Fig S1**). Detailed dependence of the probability of progression on model parameters and analytical treatment of the model will be published elsewhere.

Importantly, simulations show that hyperplasias, which progress to clinical size (i.e. $10^6$ cells, 15-20 drivers) accumulate many deleterious passengers. This evasion of purifying selection and fixation of deleterious passengers in an unexpected result that is in not programmed into the model. Although the exact number of accumulated passengers depends on the mutation rate and deleterious effect of individual passengers (**Fig. 1C**), hundreds to thousands of deleterious passengers are obtained for a broad range of parameters. Since the target size for passengers in simulations was chosen to represent non-synonymous and expressed substitution (see above), the number of passengers observed in simulations can be roughly compared to the numbers of non-synonymous substitutions observed in cancer genomics studies (**Table 2**), suggesting that some fraction of observed passengers can be mildly deleterious.

We notice two mechanisms of passenger fixation (**Fig 2B**): (i) steady accumulation, and (ii) a faster process of hitchhiking alongside a driver during a clonal expansion. Populations exhibit a large degree of heterogeneity in the number of passengers in each cell, which temporarily diminishes immediately after clonal expansions (**Fig. 2B**). Heterogeneity in the number of passengers throughout the population causes variation in the growth rates of single cells, which has been observed experimentally (38) and cannot be explained by models neglecting the effects of passengers.

Finally, simulations suggest mildly deleterious passengers, rather than highly deleterious or neutral, accumulate in large quantities and significantly affect cancer progression. **Fig 1D** presents the time to cancer onset and the number of accumulated deleterious passengers as a function of passenger's selection coefficient. Highly deleterious passengers are weeded out by natural selection, very rarely fix, and do not noticeably alter the course of cancer (**Fig. 1D**), consistent with previous findings (33).

Passengers of very small effect are effectively neutral and irrelevant to progression. However, passengers of intermediate deleterious effect fix in relatively high numbers, delay progression and the time to cancer onset, and cause extinction of the population (i.e. regression) in some cases. In fact, slower progression to cancer causes more mildly deleterious passengers to accumulate than neutral passengers. Selection coefficients for individual passengers as mild as $10^{-3}$, a phenotype typically undetectable in cell culture, are nevertheless critical for long-term population dynamics.

In summary, our simulations demonstrate that despite the mildly deleterious effect of individual passengers they accumulate in large numbers during neoplastic progression, making selection against them ineffective and potentially altering the course of neoplastic progression.

**Passenger mutations observed in cancer can be damaging**

Cancer genomics data provide an opportunity to test the assumptions and predictions of our model. First, we used comparative genomic tools to test whether non-synonymous passengers found in cancer are damaging or neutral to molecular function. Second, we tested whether selection has been acting against passengers and whether this selection was effective at preventing fixation or largely ineffective—as suggested by simulations. Alongside passengers, we analyzed driver mutations. This allowed us to validate the comparative genomic approach, as both the molecular phenotypes and positive selection experienced during cancer progression of drivers are well-known.

The Catalog of Somatic Mutations in Cancer (COSMIC), along with other cancer genomics consortia, have focused on identifying driver mutations (i.e. distinguishing drivers from passengers) by their recurrence in multiple patients or samples (32). Using COSMIC, we identified 4,195 missense (non-synonymous, amino acid changing) passenger mutations from a total of 116,977 mutations. A mutation was defined as a passenger, if it arose in a gene not listed in the census of possible cancer-causing genes (8). These 4,195 passenger mutations show no recurrence and are dispersed across 3,172 genes, further supporting their classification as passengers. We then contrasted these mutations with driver mutations and three reference datasets: (1) benign, common human non-synonymous SNPs, (2) simulated *de novo* mutations (randomly generated using a cancer-specific 3-parameter model; SI), and (3) damaging, pathogenic missense mutations causing Mendelian human diseases (from the Human Gene Mutation Database, HGMD).

We analyzed these mutations using two approaches: *PolyPhen2,* a tool widely used in population and medical genetics to assess the effects of mutations (16), and the overall ratio of non-synonymous to synonymous changes. Specifically, we used PolyPhen2's metric *ΔPSIC* that assays the degree of evolutionary conservation of a mutated residue (39). Mutations with high *ΔPSIC* scores are most likely to be damaging to molecular function (40); high scores result from mutations in well-conserved residues or from mutations that are not observed in related species. As expected, common SNPs are benign and exhibit small *ΔPSIC* values (**Fig. 3A**). In contrast, disease-causing mutations have damaging phenotypes and exhibit large *ΔPSIC* values. Driver mutations exhibit similarly high values of *ΔPSIC*, significantly greater than randomly generated mutations, indicating that drivers tend to occur at well-conserved loci in oncogenes and tumor suppressors. From a biochemical perspective, this result shows that, to activate an oncogene or to disable a tumor suppressor, the driver mutation must change a critical and well-conserved protein site, e.g. the active site of Ras or a DNA-binding site of p53. From an evolutionary standpoint, conservation of residues, which promote tumorigenesis when mutated suggests strong natural selection against the development of cancer. Ability of the ΔPSIC score to identify drivers as high phenotype mutations (i.e. damaging or altering molecular function) validates the use of comparative genomic scoring for somatic cancer mutations.

Most importantly, passenger mutations exhibit a distribution of ΔPSIC scores much greater than neutral mutations (**Fig. 3A**, $p<10^{-33}$). Hence, many passengers affect conserved residues and are likely damaging to molecular function. This result clearly demonstrates that passenger mutations are non-neutral. One caveat, that passengers could be damaging in the germ-line but neutral to cancer cells, is ruled out below. To ensure that our set of putative passenger mutations was not contaminated by drivers, we increased our stringency of passenger classification, but found no statistically significant change (p=0.69) in mean *ΔPSIC* (SI).

Moreover, passengers exhibit *ΔPSIC* values much lower that drivers (**Fig. 3A**), supporting the prediction of our evolutionary model that only passengers which are much weaker than drivers ($s_p<<s_d$) can accumulate in cancer (Fig **1D**). Finally, we notice that the *ΔPSIC* values of passengers are close to, but lower than, values of randomly generated mutations (**Fig. 3A**, $p<10^{-15}$). The similarity between passenger and random mutations suggests that many passenger mutations largely evade purifying selection. The small difference between passengers and random mutations, however, demonstrates some level of purifying selection acting against the most deleterious passengers. Collectively, our comparison of passengers and random mutations strongly supports our model's prediction that selection against mildly deleterious passengers is largely ineffective in neoplastic progression. Several important tests and controls presented below further support these results.

To confirm these results, we tested and ruled out several possible mechanisms by which passenger mutations that have damaging effect on protein function could have no effect on the fitness of cancerous cells. For example, our analyzed damaging passengers could affect genes that are functionally unimportant or not expressed in cancer cells. To test this, we considered only passengers in housekeeping genes that are essential and ubiquitously expressed, and find equally high *ΔPSIC* scores (**Fig 3B**). This rules out the possibility that damaging passengers are not expressed or present in non-functional genes. Another possibility is that only recessive heterozygous passengers show high *ΔPSIC* scores, and thus are not affecting cell fitness because the original allele provides sufficient functionality. This argument is ruled out by our observation of equally high *ΔPSIC* scores for homozygous passengers (**Fig 3B**). Finally, the deleterious effect of passengers can be manifested by their aggregation/misfolding potential that leads to proteotoxic/ER stress. Mutations in proteins that are not localized to the ER may have smaller proteotoxic effects. However, we observe equally high *ΔPSIC* scores for passengers affecting ER-localized proteins (**Fig 3B**). Lastly, to account for possible biases specific to COSMIC database, which curates mutations from the literature, we assayed passenger mutations identified in a study of 38 Multiple Myeloma genomes from TCGA Project (10) and found similar *ΔPSIC* values (**Fig. 3B, 'TCGA Myeloma'**). Collectively, our analyses show that signatures of damaging mutations are ubiquitous in known passengers and likely affect the fitness of cancerous cells.

As a powerful, *ΔPSIC*-independent, test we assayed for signatures of selection in driver and passenger genes by comparing the observed ratio of non-synonymous to synonymous mutations ($\omega$) to the predicted ratio using a random model of mutations and found several important results. First, to test whether passenger mutations could represent weak drivers, we assayed for signatures of selection in driver and passenger genes by comparing the observed ratio of non-synonymous to synonymous mutations to the predicted ratio using our random mutation model (**Fig S4**). We found that a few genes have many more non-synonymous mutations than expected under a neutral model of evolution ($\omega > 1$), which suggests that these genes are under strong positive selection; these genes were very often classified by COSMIC as drivers. Overall, most genes, and most genes classified as passengers, exhibit fewer non-synonymous mutations than expected ($\omega < 1$) suggesting these genes are under purifying selection. Similar to our *ΔPSIC* scores, the number of observed genes with $\omega < 1$ was only moderately lower than expected from our neutral model (yet still statistically significant; $p < 0.01$), further supporting the conclusion that purifying selection against deleterious passengers in neoplastic progression is ineffective.

While our genomic analysis of passenger mutations focused on missense substitutions, our model is generalizable to all inheritable genetic and epigenetic alterations, including those that are present at low frequency in the cancer population. Indeed, the length distribution of Somatic Copy Number Alterations in cancer suggests these alterations are under purifying selection as well (41). Hence, the total load of accumulated deleterious passengers in cancer may be greater than those detected by current cancer sequencing.

**The load of accumulated passenger mutations can be exploited for cancer treatment**.

Using our evolutionary model, we probed how cancers that accumulated passenger alterations respond to passenger-centric treatments. We tested two strategies: 1) increasing the overall rate of mutation ($\mu$), thus increasing the rate of passenger accumulation, and 2) magnifying the deleterious effect of passengers ($s_p$). By applying these treatments to simulated tumors we find that both treatment strategies lead to a reduction in cancer size (**Fig. 4**). Mutagenic strategies, however, require a much more severe increase (~50 fold) in the mutation rate to succeed (**Fig. 4A**), whereas a five-fold increase in $s_p$ suffices. Even with large increases in the mutation rate, the probability of 5-year relapse since treatment initiation is considerable (**Fig. 4B**). This behavior resembles patient responses to existing chemotherapeutic agents that elevate mutation rates (doxorubicin, platinum compounds, alkylating agent, etc). Perhaps, clinicians already exploit the mutagenic load of deleterious passengers by using these agents. If so, then the successes and limitations of these treatments, including relapses, may be better understood in light of the mutagenic load created by deleterious passengers.

In contrast, modest increases in $s_p$ lead to a rapid population meltdown and a low probability of 5-year relapse (**Fig. 4B**). In practice, increasing the deleteriousness of non-recurrent passenger mutations could be achieved by inhibiting cellular mechanisms that buffer against the effects of mutations (15). For example, the proteotoxic effect of passengers could be increased by targeting chaperones, proteosomes, or other components of UPR pathways (42), by elevating ER stress (43), or by stimulating misfolding of proteins destabilized by passengers using hyperthermia (44). These passenger-mediated therapies should specifically affect mutation laden cancer cells since somatic mutations are rare in normal tissues (10).

Several experiments support this clinical strategy of exacerbating the proteotoxic effects of passengers. First, chaperone proteins were found to be widely expressed in cancer and indicative of poor prognosis (45). Knockdown of *HSP1,* the master chaperone regulator, can prevent tumorgenesis in mice (42). While these other studies emphasized the role of chaperones in their ability to stabilize specific oncogenic pathways, our framework suggests cancers use chaperones to buffer the effects of passenger alterations. In our paradigm, cancer's sensitivity to chaperone inhibition results from the unleashed effects of accumulated passengers, possibly via proteotoxic/ER stress. Not only damaging missense mutations, but also chromosomal imbalances can lead to proteotoxic stress (46). Recent evidence suggests aneuploidy cancer cells rely on the UPR for survival (43). Lastly, very high levels of DNA damage and chromosomal instability correlate with positive clinical outcomes (40, 47). This last observation is inconsistent with the classical paradigm of cancer, but consistent with our framework.

Evolutionary models of cancer progression and cancer genomic studies have focused primarily on the genetic alterations required for neoplastic progression with little attention to an overwhelming majority of potentially harmful passenger alterations that arise along the way. To the best of our knowledge, this "dark matter" of cancer genomes has not been explored before. We created an evolutionary model of cancer progression that captures many key features of the process, including: a variable population size that responds to new mutations, stochastic rates of cell division and death, and, most relevant to this study, both driver and non-neutral passenger mutations. Our evolutionary model clearly demonstrates that deleterious passengers can accumulate in cancer, while our genomic analysis suggests that deleterious passengers do accumulate in cancer.

Lastly, a model that accounts for both drivers and deleterious passengers reproduces many observed phenomena in cancer, including: (i) slow initial and rapid late growth, (ii) rapid clonal expansion, (iii) dormancy and spontaneous regression, and (iv) remission upon mutagenic chemotherapy followed by relapse (35) (**Table 3**). Excluding clonal expansion (19, 20, 24) and slow initial growth (9), these effects have not been observed in previous models. Since these phenomena were not pre-programed into the model, this suggests that the key assumptions of our model are relevant to cancer progression and that the deleterious effects of passengers may, in part, explain several known properties of cancers.

## Materials and Methods

Each cell in our populations is fully described by their number of drivers $d$, and passengers $p$. Birth, death events were modeled using an implementation of the Next Reaction, a Gillespie Algorithm that orders events using a Heap Queue. All cancer mutations were collected from the ongoing COSMIC database. Common SNPs and disease causing mutations were obtained for validation of *POLYPHEN2* (3). See SI for details.

## Acknowledgments

This work was supported by the NIH/NCI Physical Sciences Oncology Center at MIT (U54CA143874). The authors would like to thank M. Imakaev, G. Fudenberg and K. Korolev for many productive discussions and commenting the manuscript.

# Figures and Tables

## Figure 1

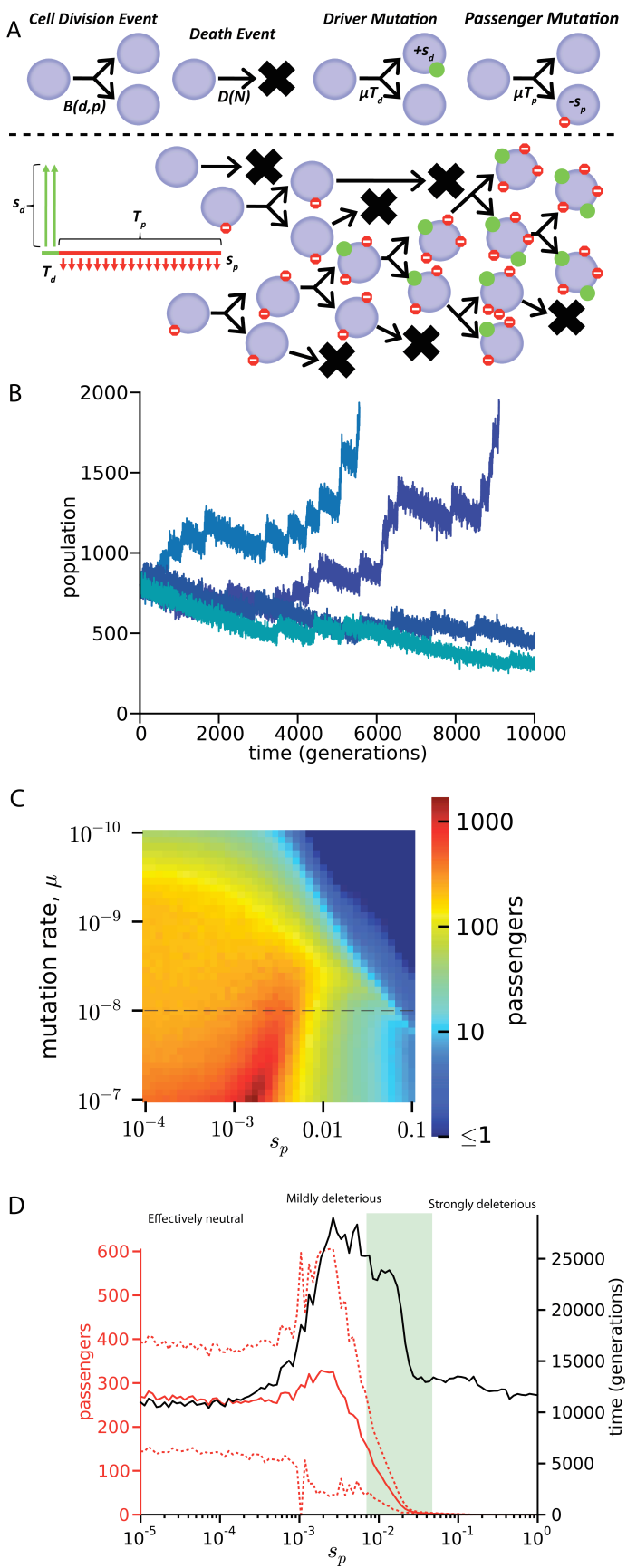

**Figure 1. Dynamics of neoplastic progression subject to driver and passenger mutations. (A)** Evolutionary model of neoplastic progression incorporating passenger mutations. Individual cells within a hyperplasia can stochastically undergo cell division (birth) or necrosis/apoptosis (death) and may acquire new driver mutations (green dot) and passenger mutations (red dot, see text for rates). Each driver mutation increases the birth rate by $s_d$, while each passenger decrease it by $s_p$. Passengers are acquired much more frequently than drivers ($T_p>>T_d$) yet have much smaller effect $s_p<<s_d$. During progression, drivers lead to clonal expansion, caring along any passengers already present in the clone. **(B)** Examples of simulated progression: despite identical parameters, they exhibit markedly different behavior, sometimes regressing to extinction or having long periods of dormancy. Steps up correspond to acquisition of a new driver. Gradual accumulation of passengers leads to slow decline. **(C)** The number of accumulated passengers depends on the mutation rate and passenger strength in a non-montonic form. **(D)** The mean number of accumulated non-synonymous passengers (solid red, dotted red = ±1 standard deviation) and the time till cancer develops (solid black line) as a function of passenger fitness effect $s_p$. Passengers of intermediate effect slow down cancer development and accumulate in larger amounts. Experimentally observed fitness effects of random point mutations in YFP in yeast ranged from 0.007 to 0.028 (sea Green).

# Figure 2

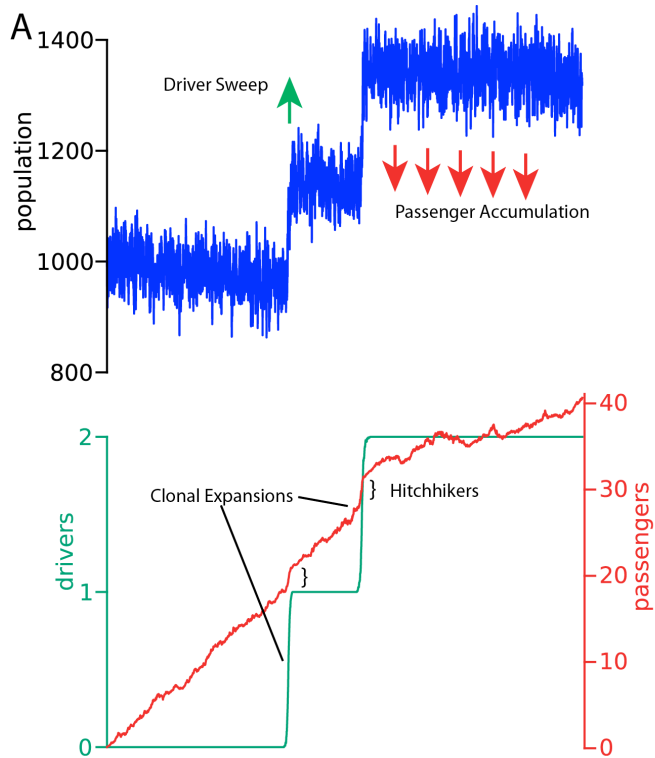

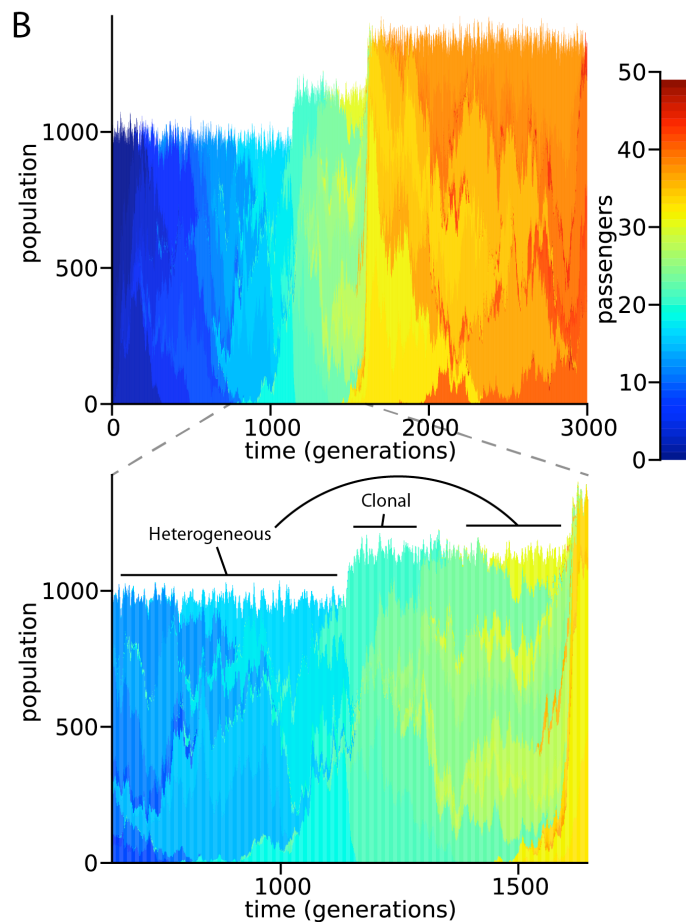

**Figure 2. Passengers accumulate gradually, via hitchhiking, and in subclonal populations. (A)** Populations grow via abrupt blooms resulting from a clonal expansion upon acquisition of a new driver mutation. In-between, the population slowly declines from gradual passenger accumulation. During clonal expansion, passengers rapidly fixate by hitchhiking. **(B)** Dynamics of sub-clones that have a different number of passenger mutations (shown by color). In between driver events, the population becomes more heterogeneous. Acquisition of a new driver leads to rapid clonal expansions and temporal loss of the heterogeneity, as all cells descend from the new fittest clone.

**Figure 3**

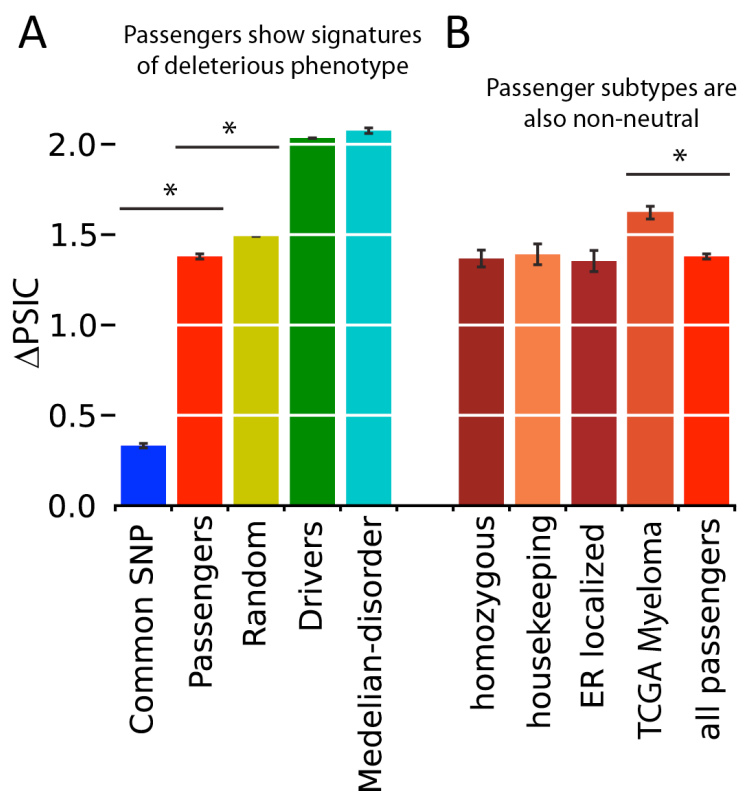

**Figure 3. Characterization of missense mutations present in cancer sequencing data.** Mutations are assayed using PolyPhern 2.0 *ΔPSIC* score, which measures the degree of evolutionary conservation of a mutated residue. Mutations with high *ΔPSIC* scores are most likely to be damaging (40). **(A)** *ΔPSIC* for passenger and driver mutations alongside three reference datasets: common (benign) SNPs, pathogenic Mendelian disease causing mutations*,* and randomly generated *in silico* mutations that experienced no selection. Driver mutations phenotypes were similar to disease causing mutations. Passengers have large *ΔPSIC* and are non-benign (compare to SNPs). Similar values for passengers and random mutations suggest that passengers largely evaded purifying selection. **(B)** Deleterious phenotypes were observed in all subsets of passengers studied, suggesting these genetic signatures cannot be explained by recessive phenotypes, or by lack of expression in somatic tissue, or any biases introduced by COSMIC's literature curation.  * *p < 0.01*.

**Figure 4**

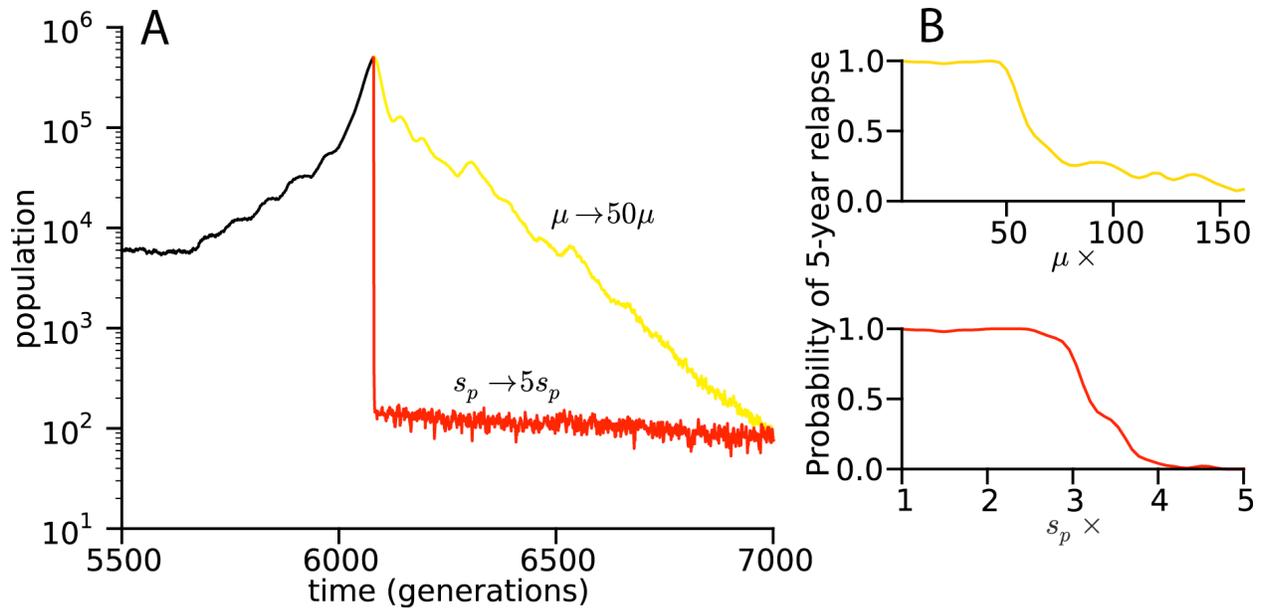

**Figure 4. Deleterious passengers can be exploited for treatment. (A)** Simulation of treatment: Populations that grow to macroscopic size can be treated by increasing the mutation rate (yellow) or deleterious effect of passengers (red). Both strategies induce cancer reduction. To induce meltdown, much greater increases in mutation rate are required as compared to increases in passenger deleteriousness. **(B)** Probability of five-year relapse after treatment as a function of increase in the mutation rate (yellow) or deleterious effect of passengers (red). Much smaller increases in the deleterious effect of passengers are sufficient to prevent the five-year relapse.

| Parameter | Symbol | Estimate | Range | Citation |
|---|---|---|---|---|
| Mutation rate | $\mu$ | $10^{-8}$ | $10^{-10}$-$10^{-7}$ | (35) |
| Driver Loci | $T_d$ | 700 | 70-7,000 | (4, 8) |
| Passenger Loci | $T_p$ | $5 \times 10^6$ | $5 \times 10^5$-$5 \times 10^7$ | (16, 33) |
| Driver strength | $s_d$ | 0.1 | 0.001-1 | (3, 20) |
| Passenger strength | $s_p$ | 0.001 | $10^{-4}$-$10^{-1}$ | (15) |
| Carrying Capacity | $K$ | 1000* | 100-10,000 | (36) |

**Table 1. Parameters of model and estimated range.** Our model contains 5 independent variables that were estimated from the literate and explored across a range of values ($\mu T_d$ and $\mu T_p$ can be abstracted to a genome-wide driver and passenger mutation rate). *Estimated from labeled populations in mice colonic crypts 2 weeks after an induced initiating *APC* deletion.

| Tumor(s) | Genes sequenced | protein coding mutations | putative driver mutations† | Reference |
|---|---|---|---|---|
| 11 Breast Cancers | 55% | 115.4 ± 53.2 | 5.1 ± 3.3 | (4) |
| 10 Colon Cancers | 55% | 75.0 ± 11.7 | 4.0 ± 0.9 | (4) |
| 4 astrocytomas, grade IV | 81% | 206 ± 343 | 5.5 ± 7.0 | (48) |
| acute myeloid leukemia | 85% | 10 | 2 | (49) |
| 26 melanomas | 85% | 366 ± 283 | 4.0 ± 4.4 | (1, 34) |
| small-cell lung cancer | 86% | 100 | 4 | (2) |

**Table 2. Accumulation of passengers coincides with theoretical results.** The mean and standard deviation of the accrued number of drivers and passengers is shown for various cancer types. In most tumors, hundreds of protein coding mutations accrue and only a few are putative drivers. These values are approximately consistent with values obtained in our model. The number of passengers varied significantly between and within tumor types, underlying the need for developing plastic models of cancer progression. The effects of deleterious passengers may be most pronounced in carcinomas since these tumors tend to have more mutations

*Based on fraction of protein coding genome sequenced assuming a complete human genome of 22,287 genes (50). These estimates, in all likelihood, underestimate the number of passengers as many are misread as sequencing errors, and no re-arrangements, copy-number alterations, changes in methylation state, or alterations outside of protein coding regions were considered. Approximately 76%(2) to 88%(1) of all substitutions are detected (False negative rate), while only an estimated 25% of indels are detected. Approximately, 97% of identified mutations are genuine (True positive rate).

† Number of protein coding mutations located in genes believed to be relevant to the cancer phenotype (8).

| Phenomenon observed in our model | Experimental Observation |
|---|---|
| **Clonal expansion, delayed growth, and extinction** | (37, 51) |
| **More mutations accumulate with high mutation rate** | (35) |
| **Approximately 50-300 total mutations acquired under realistic parameters** | (1, 2) |
| **Tumors have a large degree of heterogeneity, yet driver mutations fixate clonally** | (49, 52, 53) |
| **Mutagenic therapies often relapse after a period of remission** | (47) |

**Table 3. The deleterious passenger model reproduces many properties of cancer.** Many of the above phenomena would not be observed in our model without the inclusion of deleterious passengers. None of the above phenomena were pre-programmed into the model of neoplastic growth (i.e. population size was not fixed, nor was the number of mutations).

# The impact of deleterious passenger mutations on cancer progression

Christopher D McFarland, Gregory V Kryukov, Shamil Sunyaev, and Leonid A Mirny

## Supplemental Text and Figures

**Methods**

**Design of our model for neoplastic growth.** Cell in our populations were fully described by their number of drivers *d* and passengers *p*. Birth and death events were modeled using an implementation of the Next Reaction (1), a Gillespie Algorithm that orders events using a Heap Queue. Source code can be downloaded at https://bitbucket.org/mirnylab. Generation time in our model was defined as $1/<B(d,p)_N>$. All mutation events occurred during cell division. Dynamics should not change significantly if mutations were to occur at constant time. Because $\mu T_p$ exceeds 1 for large mutation rates, each daughter cell acquires a Poisson-distributed, pseudo-random number of new passenger mutations from its parent, with mean $\mu T_p$.

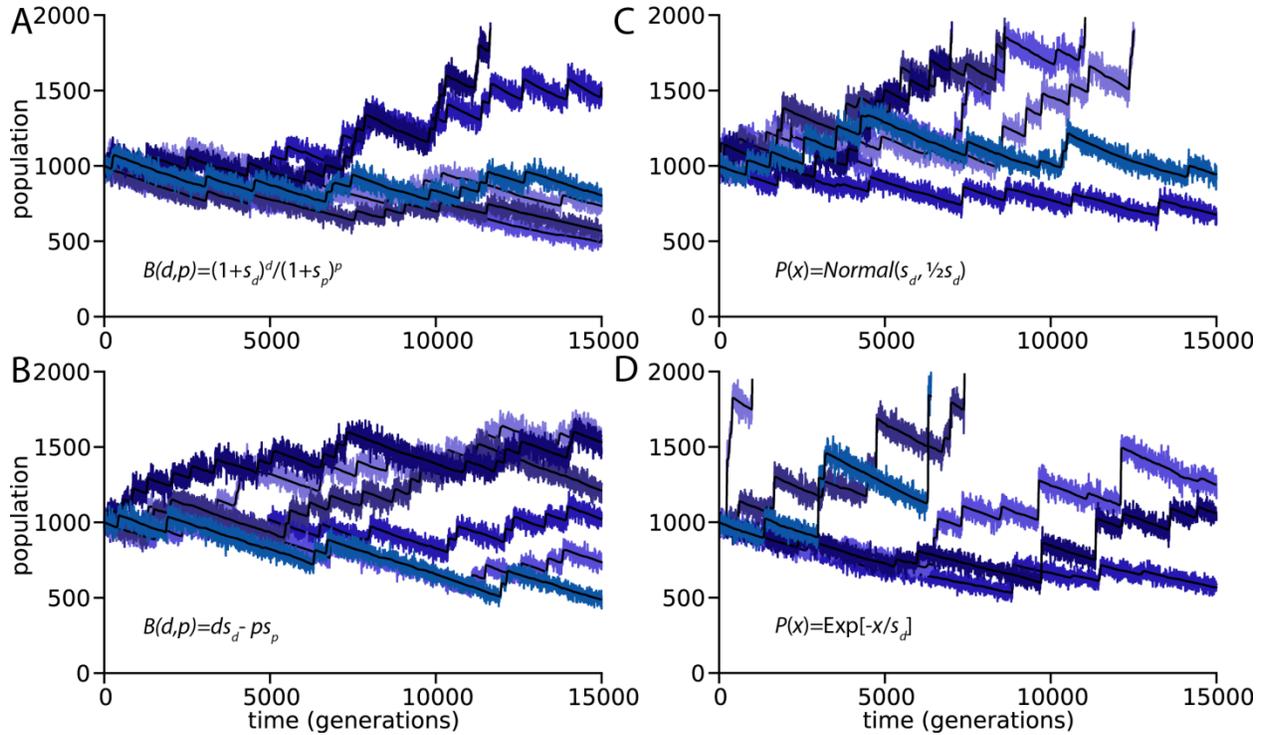

**FIGURE S1. Qualitative behavior of model is invariant to shape of birth and death functions.** Six trajectories (blue shades) with rates[1] $B(d,p) = \frac{(1+s_d)^d}{(1+s_p)^p}$ and $D(N) = \frac{N}{K}$ (A) appear qualitatively similar to simulations with rates with rates $B(d,p) = 1 + s_d d - s_p p$ and $D(N) = \frac{N}{K}$ (B). They are also mathematically equivalent to first order expansion of (d,p). Black lines represent $N$: $D(N) = B(d,p)$. The strong overlap of this black line with the observed population size indicates that birth and death are balanced throughout progression. For this reason, a model where mutations alter death rates, rather than birth rates, would be equivalent to our model[2].

Trajectories where driver loci are attributed fitness advantage by sampling from a Gaussian distribution (C) or Exponential distribution (D) appear qualitatively similar to trajectories with fixed fitness effects. All trajectories have a mean $s_d$ of 0.1.

---

[1] This functional form of *B(d,p)* is exactly equivalent to $(1+s_d)^d (1-s_p)^p$ by rescaling $s_p$.

[2] For example, suppose $B'(p) = (1+s_p)^{-p}$ and $D'(N,d) = N(1+s_d)^{-d} / K$. If $B'(p) \approx D'(N,d)$, then

$$(1+s_p)^{-p} = N(1+s_d)^{-d} / K$$

$$\frac{(1+s_d)^d}{(1+s_p)^p} = \frac{N}{K}$$

$$B(d,p) = D(N)$$

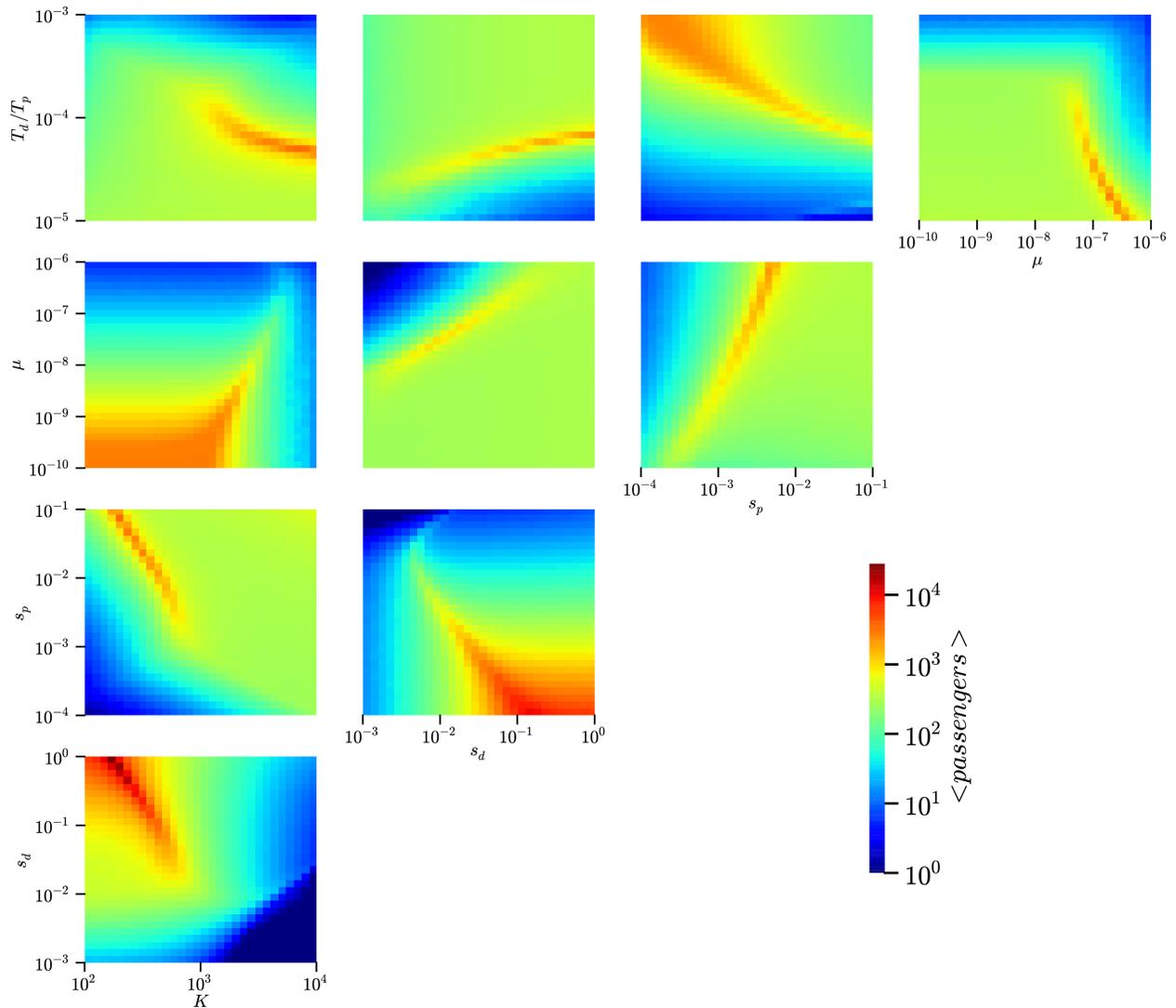

**FIGURE S2. Passengers accumulate non-monotonically across the entire parameter space.** We simulated tumor progression across all parameters. For each heatmap, the three parameters not shown on the x or y axes were set to estimated values (Table 1). For each element in the heatmap, 200 trajectories were simulated until they progressed to cancer or until 15,000 generations (~50 years), whichever was sooner. The mean number of accrued passengers is shown. Because variation in the mutation rate and $s_p$ alter the time to cancer and the probability of progression to cancer, the number of passengers depends non-monotonically on both parameters. We plan to present analytical analysis of our model's rich behavior across phase space in a future article.

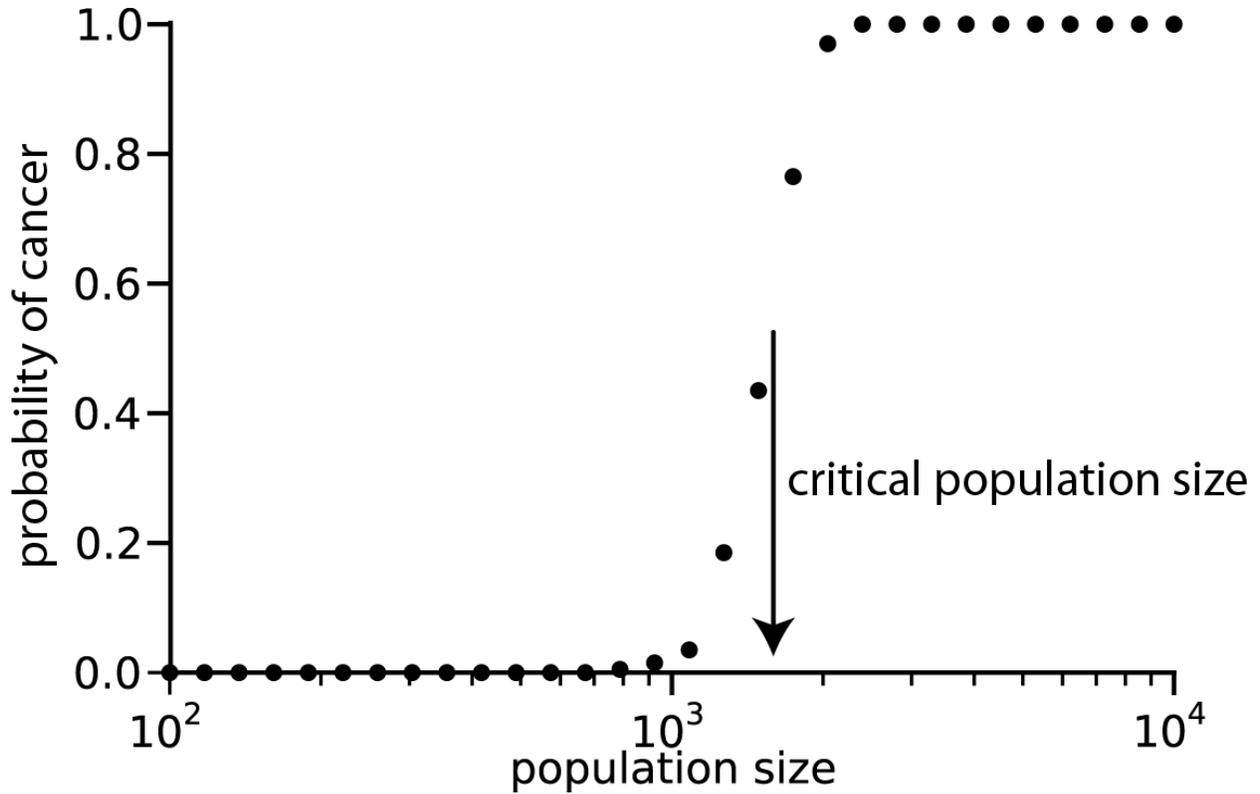

**FIGURE S3. Existence of critical population size for progression to cancer.** The probability of progressing to cancer (200 trajectory averages, $\mu = 10^{-8}$, $s_d = 0.15$, $s_p = 0.002$) exhibited a sigmoidal dependence upon the initial population size. Below this critical size, populations most often regressed from passenger accumulation.

**Critical population size**

The occurrence of a critical population size can be understood using population genetics theory and several simplifying approximations. If we assume that $B(d, p) = D(N)$ (**Fig. S1**) and consider the accumulation of advantageous drivers and deleterious passengers as independent processes, then the change in population size can be written as:

$$\frac{dN}{dt} = v_d - v_p$$

Where $v_d$ is the increase in population per unit time (velocity) due to fixation of drivers and $v_p$ is the decrease in population per unit time (velocity) due passenger fixation. New drivers and passengers arise in the population with rates $\mu T_d N$ and $\mu T_p N$ respectively. Once arose, their probability of fixation in the population is approximately $\pi_d = \frac{s_d}{1+s_d} \approx s_d$ for an advantageous driver in the absence of clonal interference, and $\pi_p \approx 1/N$ for a mildly deleterious passenger that fixates at a neutral rate. Lastly, once fixated they alter the population size by $\Delta N_d = N_{d+1} - N_d$ and $\Delta N_p$ respectively, which are found as the change in the new steady state population size upon accumulation of an additional driver or passenger: $B(d,p,N_d) = D(d,p,N_d)$, $B(d+1,p,N_{d+1}) = D(d+1,p,N_{d+1})$, yielding $\Delta N_d = N s_d$ and $\Delta N_p = N s_p$. Putting these terms together we obtain:

$$v_d = \mu T_d N \cdot s_d \cdot N s_d = \mu T_d s_d^2 N^2$$

$$v_p = \mu T_p N \cdot \frac{1}{N} \cdot N s_p = \mu T_p s_p N$$

$$\frac{dN}{dt} = \mu T_d s_d^2 N^2 - \mu T_p s_p N$$

This differential equation clearly has an unstable fixed point:

$$N_{crtical} = \frac{T_p s_p}{T_d s_d^2},$$

which roughly corresponds to the observed critical population size.

Once the population exceeds $N_{crtical}$, drivers start to have larger effect than passengers and the population expands faster than exponentially. Populations smaller than $N_{crtical}$ are more likely to go extinct (regress). To the first approximation, neoplastic development can be thought of as a stochastic process of barrier crossing, akin to many processes in chemical kinetics, where *N* is a reaction coordinate. If all passengers are neutral ($s_p$=0) the barrier is absent and progression growth faster than exponentially. The location of the barrier is set by $N_{crtical}$, which depends upon the ratio of the target sizes $T_p/T_d$ and selection coefficients $s_p/s_d^2$ of drivers and passengesrs. A large difference in the target sizes for passengers and drivers $T_p/T_d \approx 10^3 - 10^5$ (see main text) and a smaller difference in selection coefficients $s_p/s_d^2 \approx 10^{-3} - 10^0$ renders larger $N_{crtical}$ and a smaller probability of progression to cancer.

**Analysis of somatic mutations in cancer.** All cancer mutations were collected from the ongoing COSMIC database at http://www.sanger.ac.uk/genetics/CGP/cosmic/ (2). Common SNPs and disease causing mutations were obtained for validation of *POLYPHEN2* (3). As previously mentioned, drivers and passengers were categorized based on their presence in genes on or off the list of putative cancer-causing genes, updated regularly by the Sanger Institute (4). In our more stringent classification of passenger mutations, we discarded: 1) all passengers in genes that harbored more than one passenger, 2) passengers in any genes where $\omega > 1$ (see Figure S4), and 3) passengers that were not confirmed somatic mutations in the COSMIC dataset (only 29.4% of mutations in the database were confirmed by follow-up Sanger sequencing). Mean *ΔPSIC* for this stringent set of passengers did not different significantly (p <0.69) from our original set, so it was not used for further controls as it greatly reduced sample size.

To stratify passengers into various subsets, we used several resources. 372 passenger mutations were classified as 'Homozygous' by COSMIC, presumably due to some kind of Loss of Heterozygosity event. "Housekeeping" genes, were 195 genes with passenger mutations and with one-to-one orthologs in *S. cerevisiae,* identified using InParanoid (5). These genes are well expressed in humans, so we believe it is highly likely that they are expressed in cancer. Because mutations in the COSMIC database come from a variety of literature sources (which often lack direct expression data), we could not directly normalize mutations in our dataset by their expression levels. 209 genes with passenger mutations produce proteins that localize to the Endoplasmic Reticulum (ER), which we identified using the GO cellular-component category GO:0005783 (6) . 881 non-COSMIC, non-synonymous passenger mutations were obtained from The Cancer Genome Atlas' analysis of 38 Multiple Myeloma genomes (7). This subset was used as a control to ensure that any biases, which COSMIC may introduce via literature curation, did not account for our observed scores.

To parameterize our random model of pan-cancer mutations, we collected all 1,128 synonymous mutations present in COSMIC at the time of this study. Given our sample size, we parameterized our model to account for 3 types of point mutations: transversions, CpG to TpG transitions, and all other transitions, as these 3 processes seemed to explain observed mutation patterns best.  Because some genes in COSMIC, like *KRAS* or *TP53*, are sequences more often than others, we normalized both our estimated parameters and simulated mutations by the frequency with which each gene was sequenced. Hence, the frequency of mutations $f_i$ for all three mutational processes (*transversions, CpG transitions, other transitions*) were estimated as follows:

$$f_i = \sum_j^{all\ genes} w_j \frac{O_{ij}}{P_{ij}}$$

Where $O_{ij}$ is the number of observed synonymous mutations belonging to mutational class *i,* for a particular gene *j*, $P_{ij}$ is the number of possible unique synonymous mutations for the class *I* of gene *j*, and $w_j$ is the fractional of cancer sequences reported in COSMIC that belong to gene *j*. This model explained the observed patterns of non-synonymous mutations with greater log-likelihood than two-paramter models, as well more sophisticated models developed for human germ-line mutations (8). Using this model, random mutations were drawn with probability $f_i w_j$ from the set of all possible genome-wide, non-synonymous mutations. These randomly-generated mutations were not only used as a null model for evolutionary conservation, but also as a neutral null-model to test for signatures of positive and negative selection in cancer genomes.

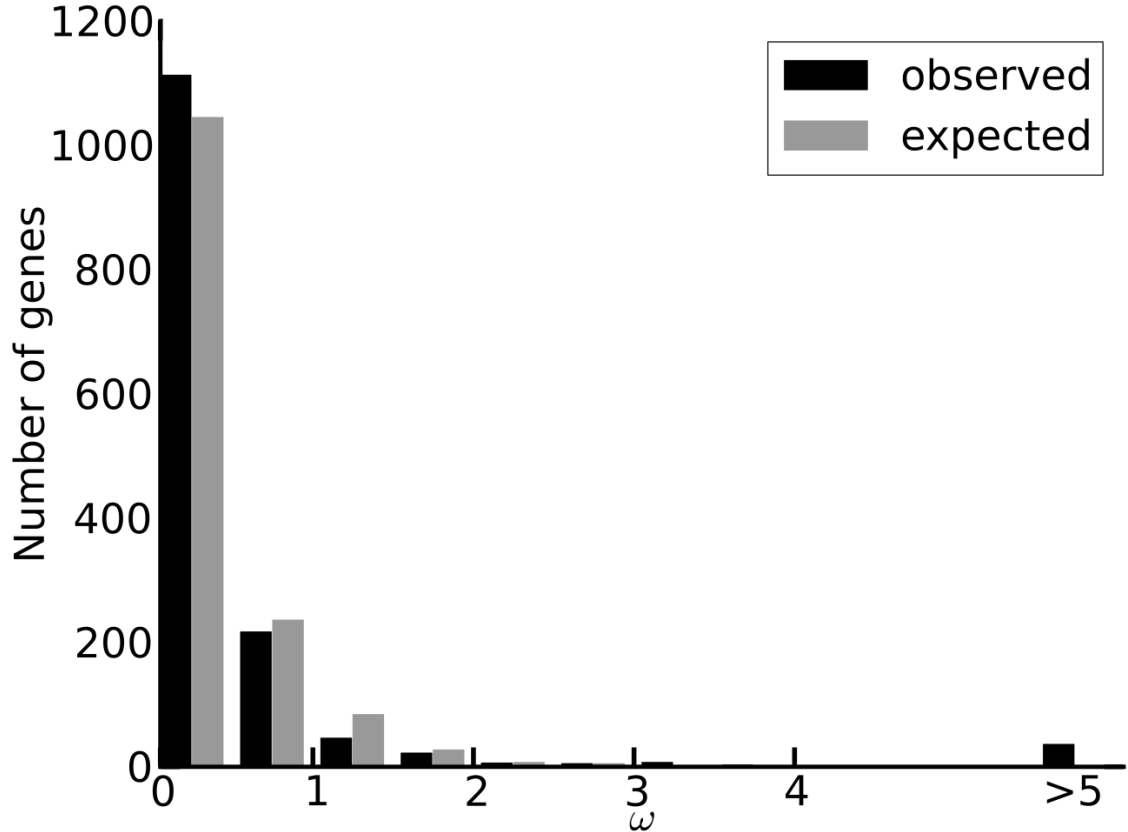

**Figure S4. Cancer mutations show evidence of positive and negative selection.** A histogram of the number of genes under positive selection (*ω* > 1) and negative selection (*ω* < 1). Because of the small number of observed mutations within each genes, a significant number of genes under neutral evolution are expected to have values of *ω* outside of 1 simply by chance (grey); nevertheless, the observed distribution has an inordinate number of genes with *ω* below 1 as well as an inordinate number above 1. This suggests that there exist both genes under purifying selection as well as genes under strong positive selection. Because many publications do not report synonymous mutations, we suspect that the true extent of purifying selection in cancer may be greater than suggested by our data from the COSMIC database.

$$\omega = \frac{O_{non-synonymous} / O_{synonymous}}{E_{non-synonymous} / E_{synonymous}}$$ Where *O* represents observed mutations in a gene and *E* represents the expected mutations in the gene using our 3-paramter random model of mutagenesis, assuming the gene experiences neutral evolution.